\documentclass{article}

\usepackage{arxiv}

\usepackage[utf8]{inputenc}
\usepackage{graphicx}
\usepackage[utf8]{inputenc} 
\usepackage[T1]{fontenc} 
\usepackage{comment}
\usepackage{palatino} 
\usepackage{amsthm, amsmath, amssymb,bm} 
\usepackage[normalem]{ulem}
\usepackage[export]{adjustbox}
\useunder{\uline}{\ul}{}

\usepackage[autostyle=true]{csquotes} 
\usepackage{algorithmic}
\usepackage{algorithm}
\usepackage{csquotes}
\usepackage{csvsimple}
\usepackage{booktabs}
\usepackage{multirow}
\usepackage{wrapfig}
\usepackage{lscape}
\usepackage{rotating}
\usepackage[final]{pdfpages}
\usepackage{titlesec} 
\usepackage{indentfirst} 
\usepackage[acronym]{glossaries}
\usepackage{array}
\usepackage{diagbox}
\usepackage{slashbox}
\usepackage{amsmath}
\DeclareMathOperator*{\argmax}{argmax} 
\makeglossaries
\newacronym{gtm}{GTM}{game-theoretical model}
\newacronym{des}{DES}{discrete event simulation}
\newacronym{ems}{EMS}{emergency medical service}
\newacronym{acs}{ACS}{Acute Coronary Syndrome}
\newacronym{pci}{PCI}{percutaneous coronary intervention}
\newacronym{gt-des}{GT-DES}{coupled game-theoretical model and discrete event simulation}
\newacronym{sa}{SA}{Sensitivity Analysis}
\newacronym{kstest}{KS test}{Kolmogorov-Smirnov test}
\newacronym{kde}{KDE}{Kernel Density Estimation}
\newacronym{ia}{IA}{inter arrivals}
\graphicspath{ {./images/} }

\title{Modelling Hospital Strategies in City-Scale Ambulance Dispatching} 

\author{
  Xinyu Fu \\
  Eindhoven University of Technology \\
  Eindhoven, The Netherlands \\[3pt]
  ITMO University \\
  Saint Petersburg, Russia \\
  \texttt{x.fu@tue.nl} \\
  \And
  Valeria Krzhizhanovskaya \\
  University of Amsterdam \\
  Amsterdam, The Netherlands \\[3pt]
  ITMO University \\
  Saint Petersburg, Russia \\
  \texttt{v.krzhizhanovskaya@uva.nl} \\
  \And
  Alexey Yakovlev \\
  Almazov National Medical Research Centre \\
  Saint Petersburg, Russia \\
  \texttt{yakovlev\_an@almazovcentre.ru} \\
  \And
  Sergey Kovalchuk\footnotemark[1] \\
  Almazov National Medical Research Centre \\
  Saint Petersburg, Russia\\[3pt]
  ITMO University \\
  Saint Petersburg, Russia \\
  \texttt{kovalchuk@itmo.ru} \\
}

\begin{document} 
\maketitle

\renewcommand{\thefootnote}{\fnsymbol{footnote}}
\footnotetext[1]{Corresponding author}

\begin{abstract}

The optimisation in the ambulance dispatching process is significant for patients who need early treatments. However, the problem of dynamic ambulance redeployment for destination hospital selection has rarely been investigated. The paper proposes an approach to model and simulate the ambulance dispatching process in multi-agents healthcare environments of large cities. The proposed approach is based on using the coupled game-theoretic (GT) approach to identify hospital strategies (considering hospitals as players within a non-cooperative game) and performing discrete-event simulation (DES) of patient delivery and provision of healthcare services to evaluate ambulance dispatching (selection of target hospital). Assuming the collective nature of decisions on patient delivery, the approach assesses the influence of the diverse behaviours of hospitals on system performance with possible further optimisation of this performance. The approach is studied through a series of cases starting with a simplified 1D model and proceeding with a coupled 2D model and real-world application. The study considers the problem of dispatching ambulances to patients with the \acrfull{acs} directed to the \acrfull{pci} in the target hospital. A real-world case study of data from Saint Petersburg (Russia) is analysed showing the better conformity of the global characteristics (mortality rate) of the healthcare system with the proposed approach being applied to discovering the agents’ diverse behaviour.


.



\end{abstract}











\section{Introduction}

Overcrowding in hospitals or emergency departments has detrimental effects because of rising patient volumes and profound numbers of admitted patients waiting for service \cite{kamal2014addressing}. Incidents have been reported that patients with severe syndromes were suffering from prolonged waiting \cite{mccarthy2009crowding,schull2004emergency}. An essential reason behind overcrowding is the imbalanced allocation of medical resources in a city \cite{ruger2012patterns}, which led to regional overcrowding and varying health outcomes of patients among hospitals. 

\acrfull{acs} patients are suffering from overcrowding issues in metropolitan areas of large cities (here and further we consider Saint Petersburg, Russia, a city with over 5M citizens, as an example). In Saint Petersburg, hospitals overcrowding is often caused by the irregular inflow of patients and the limited number of specialists and facilities (mainly angiography devices used for the \acrshort{pci}). Consequently, the annual mortality rate of the \acrshort{acs} patients dramatically ranges from 4\% to 18\% among all hospitals in Saint Petersburg that have the \acrshort{pci} facilities in 2015. One of the possible solutions is the improvement of the citywide ambulance dispatching system by optimizing the patient's overall time spent and balancing the patient and medical resources. The ambulance dispatch process describes the process starting from patients being picked up and pre-treated by an ambulance team and ending with being transported to the destination hospital. Particularly,  the paper focuses on the optimization of destination hospital selection problems. The choice of the destination hospital is related to the patient's transportation time and operation waiting time before treatment, which indirectly and significantly affects the health outcome (particularly for the \acrshort{acs} patients). The decision of assigning patients to a destination hospital within ambulance dispatching is announced by the \acrfull{ems} but decided collectively with the receivers (i.e. potential target hospitals). Since the \acrshort{ems} plays a messenger role and its overall goal is equivalent to the system objective \cite{aboueljinane2013review}, its dynamic operation has less space to be improved. However, the receivers' (hospitals) decision-making process is individual-wise and does not have to be globally optimal, a landing point of our model is to predict their strategies and actions.

The study focuses on discovering the diversity in hospital behaviour formed within interactive decision making considered as a non-cooperative game. The proposed approach is based on the combination of discrete event simulation (DES) and an equilibrium-based framework using game theory (GT) to simulate the patient flow and hospitals’ responding strategies in order to understand the dynamics in the ambulance dispatching process. The work is based on the idea \cite{kovalchuk2018iccs} showing that the global optimal behaviour of healthcare agents may differ from the equilibrated and real behaviours. This work develops this idea with a multi-agent hybrid (the \acrfull{gt-des}) model and applies it to the real-world example of Saint Petersburg to evaluate the proposed approach. Also, the paper is based on the MSc thesis by Fu \cite{fu2020msc}, which can be referenced to describe extended experiments.

The remaining part of the paper is structured as follows. Section \ref{rel_works} presents a brief review of related works. Section \ref{methodology} describes the basic methodology of the study and introduces the hybrid models used to investigate the diverse behaviour of the healthcare agents. Next, Section \ref{case_study} describes experiments having an ambulance dispatching in Saint Petersburg as an example. Finally, Section \ref{discussion} and \ref{conclusion} presents discussion and concluding notes respectively.

\section{Related works}
\label{rel_works}

The study of related works focuses on using computer-based stochastic simulation in the context of the ambulance dispatching process. Stochastic simulation is defined as a computer-based simulation of a system where the variables of the system can change randomly with individual probability. The method aims to design a computational model that mimics a system’s operations in order to better understand the behaviours of the system \cite{graham2013stochastic}. As a commonly used tool in computational science and operational research, stochastic simulation has proven itself in various fields for the past two decades (e.g. in finance \cite{huynh2011stochastic}, telecommunication networks \cite{pawlikowski2002credibility}, and transportation \cite{chen2004stochastic}).

\subsection{Computer-based simulation for the \acrshort{ems}-related systems}

Particularly in the context of the \acrshort{ems}-related systems, \cite{zhen2014simulation} designed a framework to optimise the process of ambulance deployment and relocation. \cite{janovsikova2019optimization}  proposed a simulation-based approach to optimise the relocation of emergency departments, and \cite{ramirez2011design} used simulation optimisation to improve the performance of an emergency medical service. On the other hands, various tools are derived from computer-based simulation in the \acrshort{ems}-related systems. Early in the 90s, mathematical programming was commonly discussed in this area \cite{baker1989non}. Because of the increasing computational power, discrete event simulation had been a general method utilised for complex system simulation \cite{jacobson2006discrete}. Nowadays, stochastic queuing models have been widely accepted in simulating the operation of the \acrshort{ems} \cite{aboueljinane2013review}. However, in reality, \acrshort{ems}  in various regions are similar but not simple \cite{aboueljinane2013review}. An \acrshort{ems}-related system and its simulation normally have a high degree of complexity, uncertainties due to the complex process and various stakeholders (e.g. hospitals, patients, city managers, and the \acrshort{ems} itself) involved. Consequently, the related work focuses on operations, processes, and specific features of the simulated \acrshort{ems}-related systems in order to capture the scope and complexity associated with a \acrshort{ems}-related system. 

\subsection{Destination hospital selection}


When ambulance teams pick up and pre-treat patients, a destination hospital is selected by the \acrshort{ems}. Most research in this field assumes that the selection rule is static and the selection process is based on a stationary demand pattern. For instance, a widely used rule is called the principle of proximity, where ambulance teams transport patients to the nearest hospital. However, both our patient transportation data and other studies show that more than 40\% of transported patients were not treated by the nearest hospital \cite{savas1969simulation}. Reasons including the patient’s personal intentions and financial capacity, the treatment capacity of the destination hospital, the capacity of the currently available resources, and the estimated delivery time. In addition, some researchers use the historical distribution of the patients' geographic location as the main reason for selecting the destination hospital \cite{henderson2005ambulance}. Another static selection method uses a pre-defined schedule to assign a request call to a destination hospital \cite{silva2010emergency}.

\subsection{Dynamic ambulance redeployment}

Rather than statically selecting destination hospitals, ambulances normally change target hospitals based on the real-time situation. Stakeholders such as patients, hospitals, the  \acrshort{ems}, and ambulance teams change their behaviours based on their interests and social responsibility. For instance, the expected travel time may vary in rush hours or due to different path planning for an ambulance team \cite{henderson2005ambulance}. However, there are very few studies simulating the dynamic redeployment of ambulances. It is a complex issue because it includes the consideration of time-dependent behaviours and the real-time availability of medical resources. Still, it is proven that a dynamic policy performs better compared to a static policy\cite{daskin1983maximum}.

\section{Methodology: coupled dispatching model}
\label{methodology}

In this section, the methodology of the \acrfull{gt-des}) model is explained. From simplicity to complexity, the \acrshort{gt-des} model is firstly applied to one-dimensional and two-dimensional artificial scenarios. The aim is to firstly understand the general ambulance dispatching system by simulating patient flow (incl. patients’ arrival, travel, and service process). Secondly, to  predict the hospital’s action if they accept or redirect patients with respect to varying arrival and service rates.  For more simulation details, please refer to the dissertation \cite{fu2020msc}.
 
\label{conceptual_model}
\begin{figure}[!h]
 \centering
   \includegraphics[width=1\textwidth]{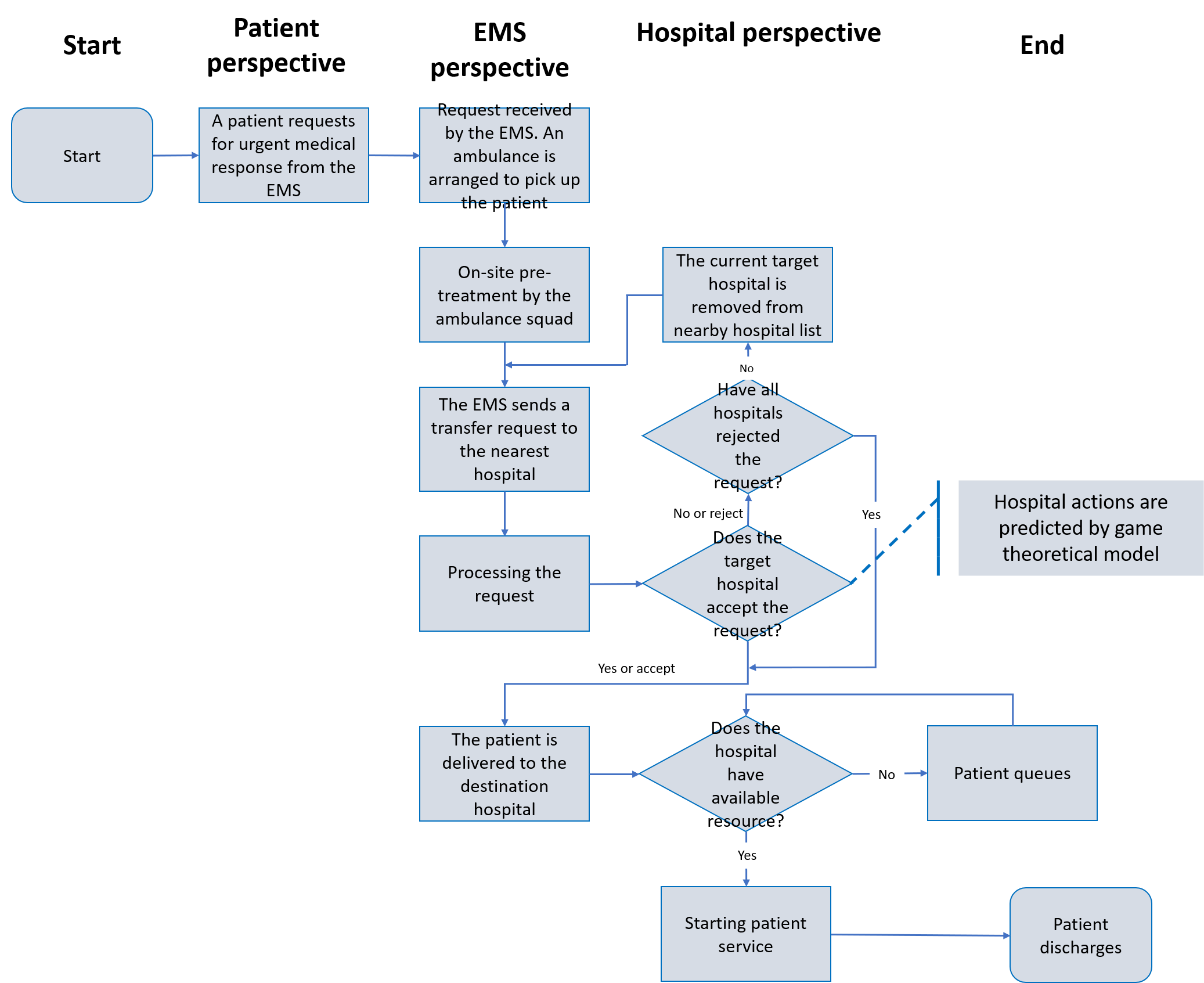}
   \caption{A general ambulance dispatching process in Saint Petersburg}
\label{fig:flowchartADM}
\end{figure}

Figure~\ref{fig:flowchartADM} conceptually describes the ambulance dispatching process from perspectives of patients, the \acrshort{ems} and hospitals (using the dispatching system in Saint Petersburg as an example). Generally, once a patient is picked up by an ambulance, the nearest hospital is chosen as prioritised destination so called target hospital. Each hospital has either "Accepting" or "Redirecting" strategy. The patient is sent to target hospital if the target hospital assumes the “Accepting” strategy. However, suppose that the target hospital’s strategy is “Redirecting”, and the current number of patients has surpassed the predefined capacity, the ambulance brings the patient to the next nearest hospital and update it as the target hospital.  This process is repeated by asking over all the nearby hospitals. If all hospitals reject the patient’s requests, the patient is sent to the most suitable hospital with the minimum time spent on travelling and serving. After patients enter the hospital, they take turns to queue, get served and discharge from the hospital. It shall be addressed that in an ambulance dispatching system, the definition of “server” is any medical resource that is realistically most crucial to patient healing, which can be medical equipment, professional doctors, and beds.
 
\subsection{Simplified 1D model}
To simulate Figure~\ref{fig:flowchartADM} in a simplest way, we designed a one-dimensional \acrshort{gt-des} model which depicts the process where an unlimited number of patients take turns to arrive and discharge from two hospitals.

\begin{figure}[!h]
\centering
\includegraphics[width=0.4\textwidth]{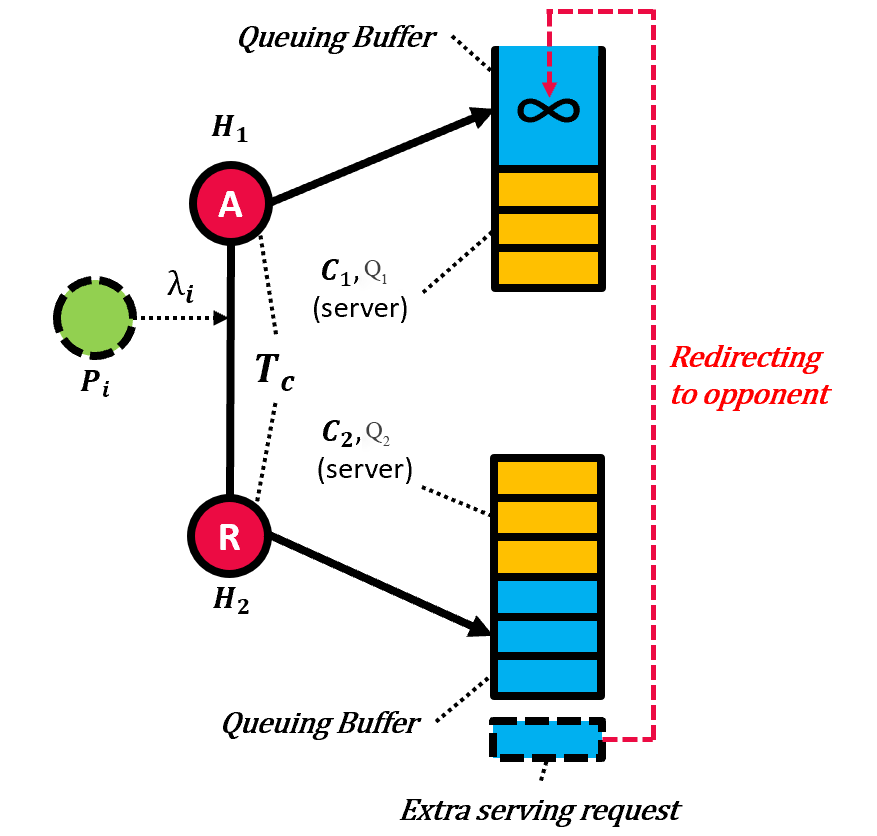}
\caption{The structure of a one-dimensional ambulance dispatching model}
\label{fig:1dModel}
\end{figure}
    
Figure~\ref{fig:1dModel} shows the structure of a 1-D queuing network model describing the process of patient arrival, travel, and service. Formally, there are $k=2$ hospitals on each side of the map. Each hospital is denoted as $H_j$, with unique id $j$ for $j=\{1,2\}$.  Each hospital $H_j$ can assume either the strategy of “Accepting” (abbr. $A$, accepting all patients) or “Redirecting” (abbr. $R$, redirecting patients to the opposite hospital if the current number of patients in the hospital exceeds its limited capacity). The capacity $N_j$, is equivalent to the sum of a limited number of servers $C_j$ (e.g. surgery equipment) and a limited or unlimited length of queuing buffer $Q_j$ (e.g the size of waiting room), which results in $N_j=C_j+Q_j$. Patient $p_i$  with unique id $i$ randomly and constantly appears on the map at a random time. First, patient $p_i$ requests a service at the nearest hospital or the second nearest hospital if the first request is rejected. The approval of the request is based on the hospital's strategy and resource availability. In case no hospital accepts patient's request, the nearest hospital has to accept the patient's request by the arrangement of the \acrshort{ems}.  Next, patients are delivered to the target hospital and get served immediately if any server is available. If none of the servers is available, the patient is queued until a server is available. The service process follows the principles of "First Come First Served" and "First In First Out". Lastly, the patient discharged after the service is completed.

To simulate the process, we implement a $M/M/C$ queuing model \cite{banks2005discrete} with arrival rate $\lambda$ and service rate $\mu$ for each server and the number of $C_j$ servers for each hospital $j$ ( we require $\rho =\frac{\lambda}{\sum^k_{j=1} c_j \mu} < 1$ otherwise the system will explode because of overcrowding). The model contains three parts which are arrival process, service process and game process. 


\textbf{Arrival Process} 
The arrival process describes the period starting from a patient being picked up by the ambulance to their arrivals at the hospital. Arrivals occur at rate $\lambda$ according to a Poisson process \cite{banks2005discrete}. Thus, an inter-arrival time is theoretically considered to have an exponential distribution with $\lambda$ as the mean arrival rate (defined as the average number of arrivals per time period, which is equivalent to the request rate in the simulation \cite{banks2005discrete}. Thus, time that elapsing between two arrivals (the \acrfull{ia}) is depicted by $$ P(IA<=T) = 1 - P(IA>t) $$

By replacing $P(IA>t)$ with its cumulative distribution function, we obtain: $$P(IA<=T) = 1-e^{-\lambda t} $$

Thus the probability that there are exactly $x$ arrivals during $t$ amount of time is $$ P(x) = \frac{e^{-\lambda t} (\lambda t)^x}{x!} $$

Since two hospitals have a of maximum four combinations of strategies (e.g. AA, AR, RA, RR, where AR means that one hospital’s strategy is “Accepting” and the another one’s strategy is “Redirecting”. This norm applies for AR, RA, and RR as well), the effective arrival rate at each hospital is updated upon each strategy combination. For more information about the analytical expression of the updated arrival rate, refer to \cite{fu2020msc}.

\textbf{Service Process} The service process begins from a patient’s service to their discharge from the hospital. A key argument to the simulated service process is the service time. Theoretically, the service time of each patient has a Poisson distribution with the service rate $\mu$ (the average number of services per unit of time per server)\cite{banks2005discrete}. For each hospital, there are $c$ servers, and if there are fewer than $c$ jobs, the remaining servers are idle. If there are more than $c$ jobs, the extra jobs queue in a buffer \cite{banks2005discrete}.

\textbf{Game Process} 
The game process describes that hospitals play a non-cooperative game against each other. For the destination selection problem, hospitals' responses decided where patients are going. The process is aimed at predicting hospitals' strategies of "Accepting" or "Redirecting" patients by assuming a stationary equilibrium (Nash Equilibrium) is formed. Nash Equilibrium is a concept derived from game theory where no player has anything to gain by changing their own strategy \cite{shubik1982game}. However, there are three assumptions in our simulation:
\begin{enumerate}
    \item the \acrshort{ems} only plays the role of a messenger by sending requests to nearby hospitals and transfer the approval or reject message back to ambulance and patients, rather than arranging it directly. A hospital has the right to accept or redirect patients based on internal information about their resource availability and empirical estimation.
    
    \item Each strategy combination needs to be simulated. However, hospitals keep their strategies unchanged for each round of simulation. 
    
    \item A hospitals’ actions are motivated by the payoff which incorporates its social responsibility, budget and resource utilisation.

\end{enumerate}


Thus, we introduce a payoff metric named "Score" which describes the performance of a hospital contributing to the global optimal. The "Score" of a hospital $H_j$ is defined as

\begin{equation}
    Score_j =\frac{n_{served}^j}{T_{total}^j}
\end{equation}

where $n_{served}^j$ is the number of patients served in hospital $j$ within the simulation time. The average total time  $T_{total}^j$ spent for each patient in hospital $j$ is the sum of average travel, queuing, and service time

\begin{equation}
    T_{total}^j=T_{travel}^j+T_{queue}^j+T_{service}^j
\end{equation}

\begin{table}[]

\centering
\caption{A payoff matrix $ \in \mathbb{R}^{2 \times 2 \times 2}$ of a non-cooperative game with three hospitals.}

\begin{tabular}{|c|c|c|c|c|c|}
\hline
                                     &            & \multicolumn{4}{c|}{\textit{Hospital 1}}                                                      \\ \hline
                                     &            & \multicolumn{2}{c|}{\textbf{A}}               & \multicolumn{2}{c|}{\textbf{R}}               \\ \hline
                                     &            & \multicolumn{2}{c|}{\textit{Hospital 2}}      & \multicolumn{2}{c|}{\textit{Hospital 2}}      \\ \hline
                                     &            & \textbf{A}            & \textbf{R}            & \textbf{A}            & \textbf{R}            \\ \hline
\multirow{2}{*}{\textit{Hospital 3}} & \textbf{A} & $u_A^1, u_A^2, u_A^3$ & $u_A^1, u_R^2, u_A^3$ & $u_R^1, u_A^2, u_A^3$ & $u_A^1, u_R^2, u_A^3$ \\ \cline{2-6} 
                                     & \textbf{R} & $u_A^1, u_A^2, u_R^3$ & $u_A^1, u_R^2, u_R^3$ & $u_R^1, u_A^2, u_R^3$ & $u_R^1, u_R^2, u_R^3$ \\ \hline
\end{tabular}

\label{fig:payoff}
\end{table}

Table \ref{fig:payoff} shows an example of the $ \mathbb{R}^{2 \times 2 \times 2}$ payoff matrix, where $\bm{N}=\{1,2,3\}$ is a finite set of three hospital agents. We define $\bm{A}=A_1 \times A_2 \times A_3 $ a set of available actions where $A_j$ is the set of actions available to agent $j$ for $j \in \{1,2,3\}$. And action profiles $\bm{a}=(a_1,a_2,a_3)$ where $a_j \in A_j = \{A,R\}$ with $A$ as "Accepting", and $R$ as "Redirecting" strategy or action.  $\bm{u}=(u^1,u^2,u^3)$ is a profile of utility function and $u_a^j$ is the utility of hospital $j$ taking action $a$ \cite{leyton2008essentials}. A example of the payoff outcome from a round of simulation is $\bm{u}=(u_a^1, u_a^2, u_a^3)$ where $a=\{A,R\}$. As a result, by applying Algorithm \ref{alg:search_ne} to search for Nash equilibrium, we derive a stable equilibrium where hospitals don't intend to change their strategies.

\begin{algorithm}[H]
\begin{algorithmic}[1]
\FOR{$a_1$ in $\{a,a^{-1}\}$}{
    \FOR{$a_2$ in $\{a,a^{-1}\}$}{
        \FOR{$a_3$ in $\{a,a^{-1}\}$}{
            \IF{$u_1^{a_1,a_2,a_3}\geq u_1^{a_1^{-1},a_2,a_3}$ 
            and $u_2^{a_1,a_2,a_3}\geq u_2^{a_1,a_2^{-1},a_3}$
            and $u_3^{a_1,a_2,a_3}\geq u_2^{a_1,a_2,a_3^{-1}}$} 
                \STATE {$a_1,a_2,a_3$ is weak Nash Equilibrium} 
            \ENDIF
        } 
        \ENDFOR
    } 
    \ENDFOR
} 
\ENDFOR
\RETURN {Nash Equilibrium from a matrix $ \in \mathbb{R}^{2 \times 2 \times 2}$}
\end{algorithmic}


    \caption{An algorithm to search for Nash Equilibrium among three players}
\label{alg:search_ne}
\end{algorithm}

To represent the system performance, we derive two long-term metrics

\begin{itemize}
    \item To describe the queuing performance, the average number of patients in queue $L_Q$ and average time of each patient spent in queue $W_Q$ are used  \cite{banks2005discrete}:
    $$L_Q=\frac{P_0a^c\rho}{c!(1-\rho)^2)}[1-\rho^{N-c}-(N-c)\rho^{N-c}(1-\rho)]$$
    
    $$W_Q=L_Q/\lambda_e$$
    \item To describe the system performance, the global time, an indicator matching up the objective of the \acrshort{ems} is used: 
    $$T_{global} = \frac{T_{total}^1\lambda_1 + T_{total}^2\lambda_2}{T_{total}^1+T_{total}^2}$$
\end{itemize}

\begin{figure}[!h]
\centering
\includegraphics[width=1\textwidth]{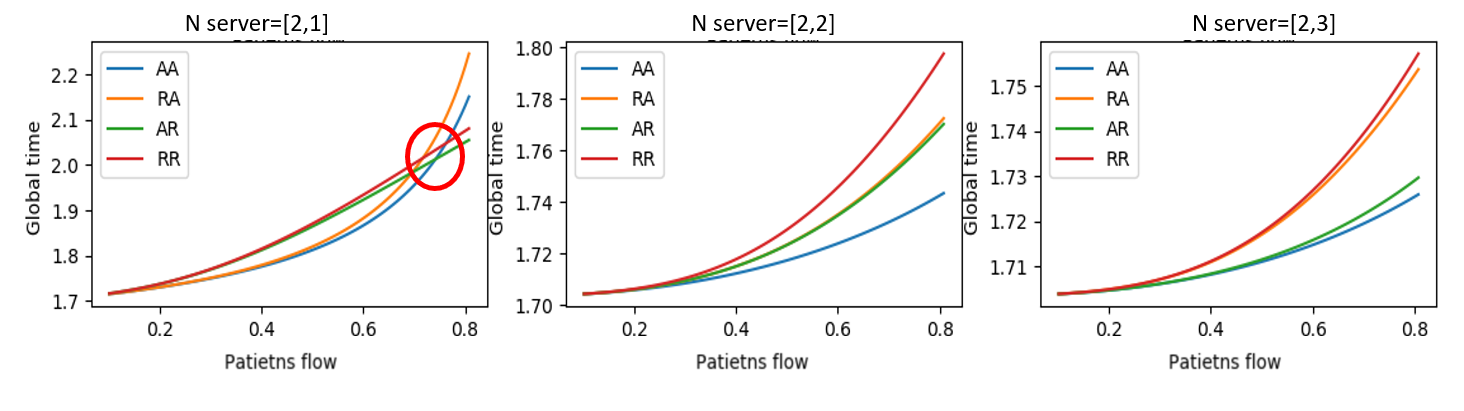}
\caption{The trends of the global time with different combinations of server capacities}
\label{fig:1dGloball}
\end{figure}

\textbf{Results of the one-dimensional experiments} 
To find the optimal strategy for the system, Figure~\ref{fig:1dGloball} presents the average global time for every patient appeared (note that the less the global time, the better for the system). It shows that the $AA$ combination of strategy is the most beneficial in most cases by comparing the global time. However, at the circled intersection, the best strategy combination switches. With the higher patient flow, optimal strategy changs from $AA$ to $RA$ and $RR$, under the conditions of non-equivalent servers $\bm{N}=\bm{\{N_1, N_2\}=\{2,1\}}$. The result illustrates the potential reason for which a hospital switches its strategy may come from server inequality and irregularities of patient flow. 

\subsection{2D model-based ambulance dispatching}
\label{2dmodel}

However, the optimal strategy combination is not always applied for a non-cooperative environment. To find out which strategy combination is realistic, we conduct a two-dimensional simulation to predict the stationary equilibrium \cite{kovalchuk2018iccs}. Thus, the two-dimensional \acrshort{gt-des} model is designed for a city environment.

\begin{figure}[!h]
\centering
\includegraphics[width=0.4\textwidth]{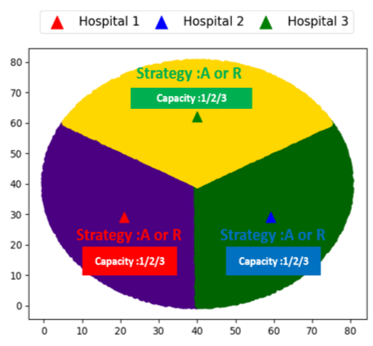}
\caption{Simulated two-dimensional map with three hospitals and corresponding capacity of servers}
\label{fig:2d_map}
\end{figure}

\textbf{Problem statement} To experiment with the two-dimensional simulation, we artificially set an environment. Figure~\ref{fig:2d_map} shows the spatial environment of the two-dimensional simulation. A simulated area has three hospitals evenly distributed in the cluster area. Shaded sectors are "service coverage" zones. Each hospital has the same Euler distance to any other hospital. Shaded sectors are the "service coverage" zones. Patients from each zone are normally delivered to the nearest hospital if the hospital accepts their request.   our question is which strategy combination is formed given different environmental factors (i.e arrival rate, service rate, travel time, resource capacity etc). 

\textbf{Sensitivity Analysis} 
To evaluate the importance of each environmental factor, we implement Sensitivity Analysis. The \acrfull{sa} is a method utilised to quantitatively evaluate how much of the effect of the output comes from each input to model \cite{saltelli2008global}. In the section, we utilised the Sobol method \cite{saltelli2004sensitivity} which is a global method that reveals the influence resulted from both individual input factors and interactions between input factors. Besides, passing the Sobol method to a Non-linear process is commonly used\cite{saltelli2004sensitivity}. In Appendix.\ref{Appendix:samplesize}, we carry out an experiment to find the size of distinct samples for the Sobol method, the result shows distinct sample size $N=2^{14}$ can satisfy achieving the range of $\pm 0.1$ for "score" with 95\% confidence. 

\begin{figure}[h!]
\centering
\begin{minipage}{1\textwidth}
\centering
\minipage{1.0\textwidth}
\centering
\includegraphics[width=0.8\textwidth]{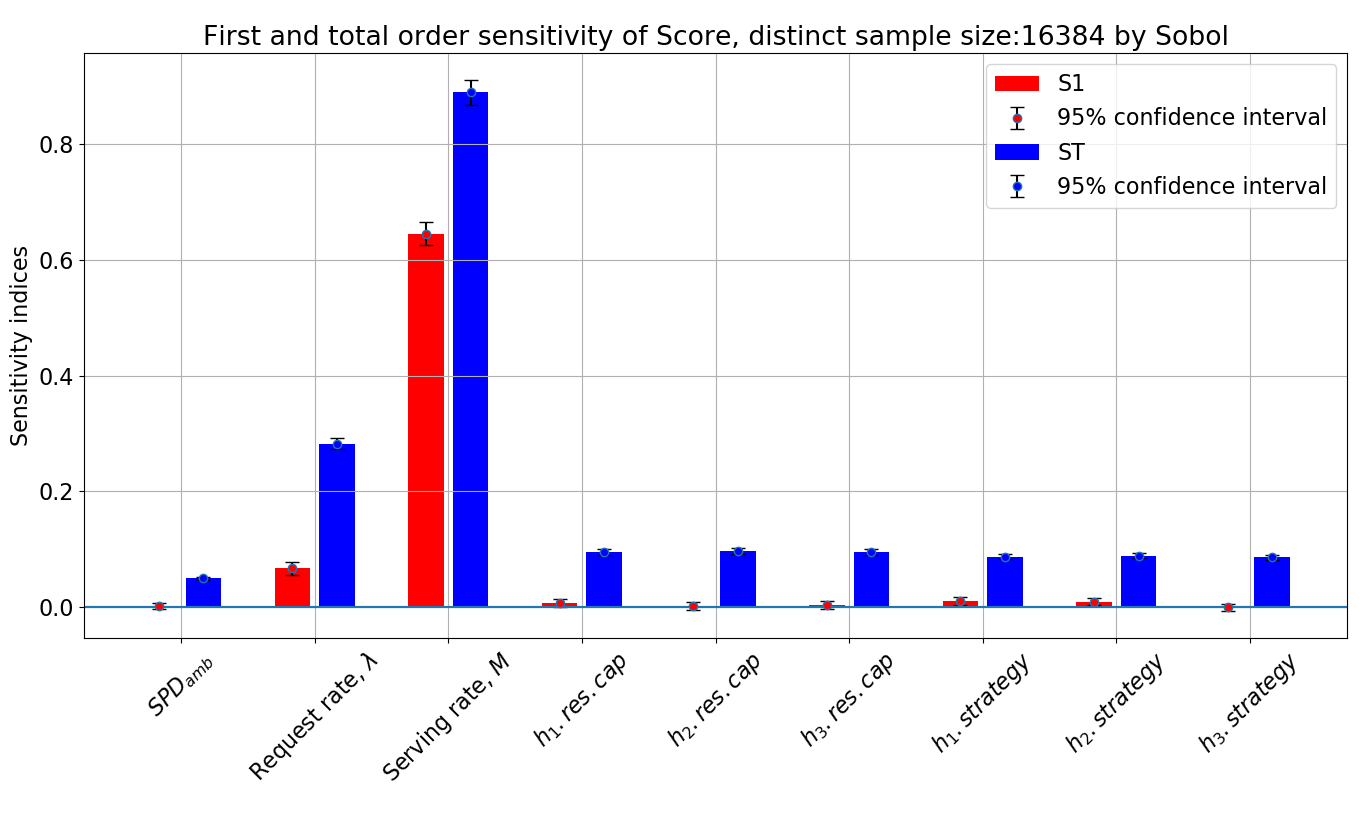}
\centering
\includegraphics[width=0.8\textwidth]{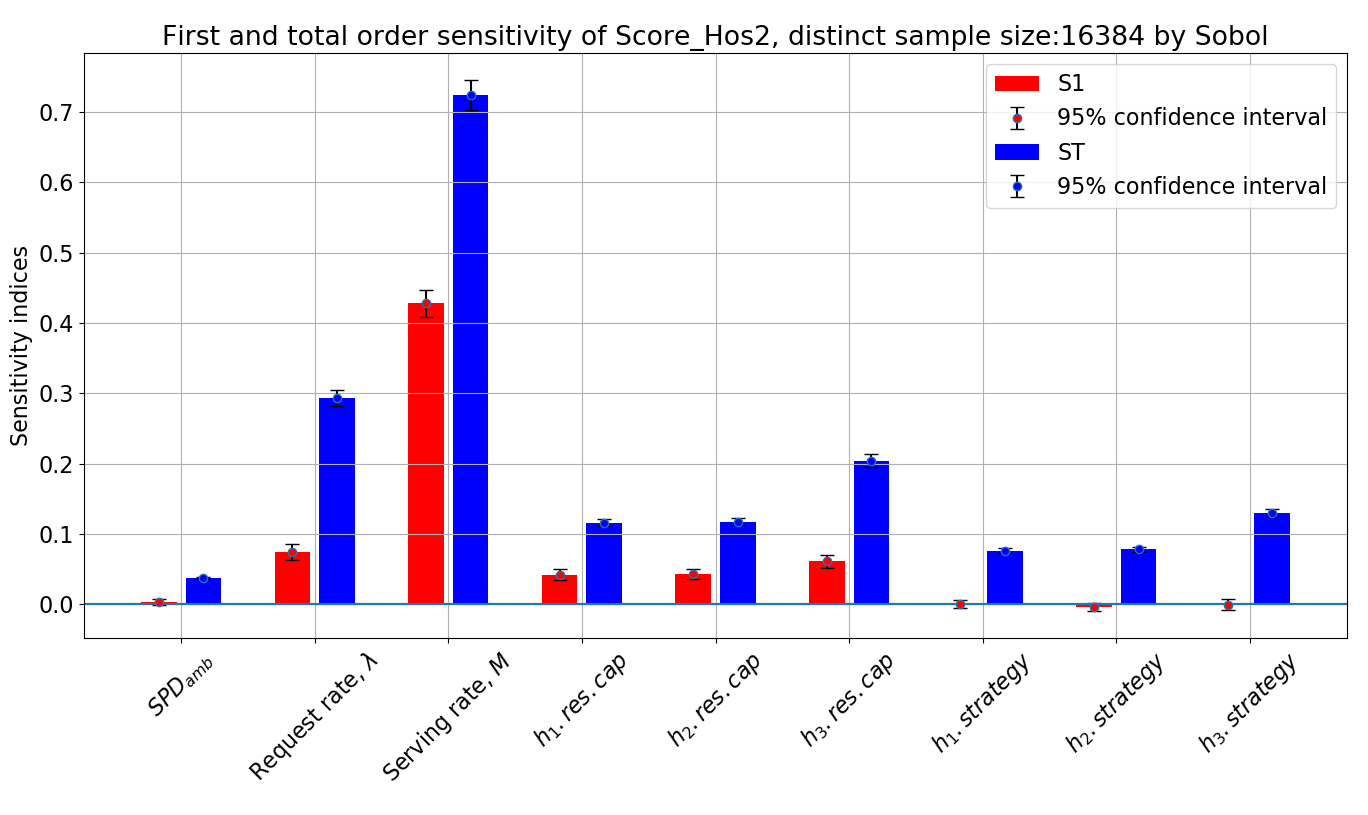}
\endminipage\hfill
\caption{The first and total order sensitivity with the payoff "Score" (top) and the "Score of hospital 2" (bottom).}
\label{fig:sa_score}
\end{minipage}\hfill
\end{figure}

As a result, Figure~\ref{fig:sa_score} shows that the arrival rate $\lambda$ (or request rate) and service rate $M$ contribute most of the sensitivity to either the system or the score of any hospital. The velocity of ambulance contributes significantly less than the other factors. This idea is used later in the case study where travelling time is not as important as door-to-balloon time. 

\begin{figure}[!ht]
\centering
\begin{minipage}{1.0\textwidth}
\centering
\includegraphics[width=0.9\textwidth]{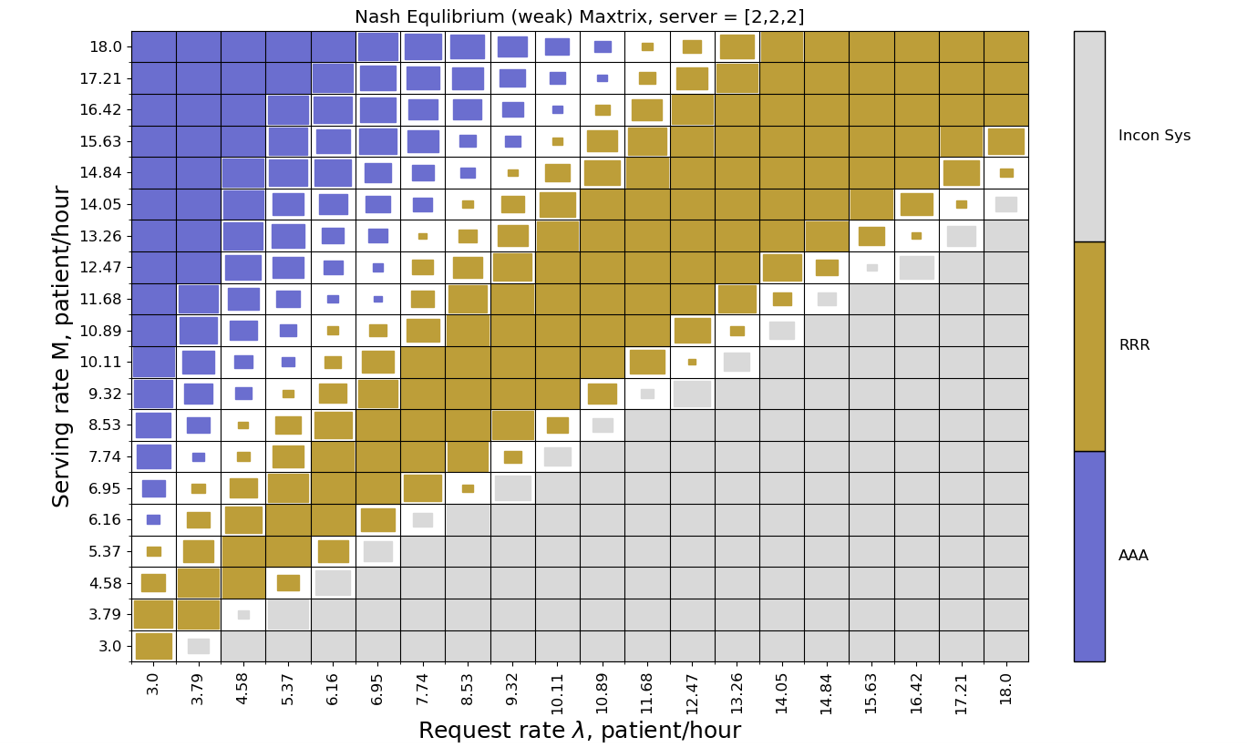}
\includegraphics[width=0.9\textwidth]{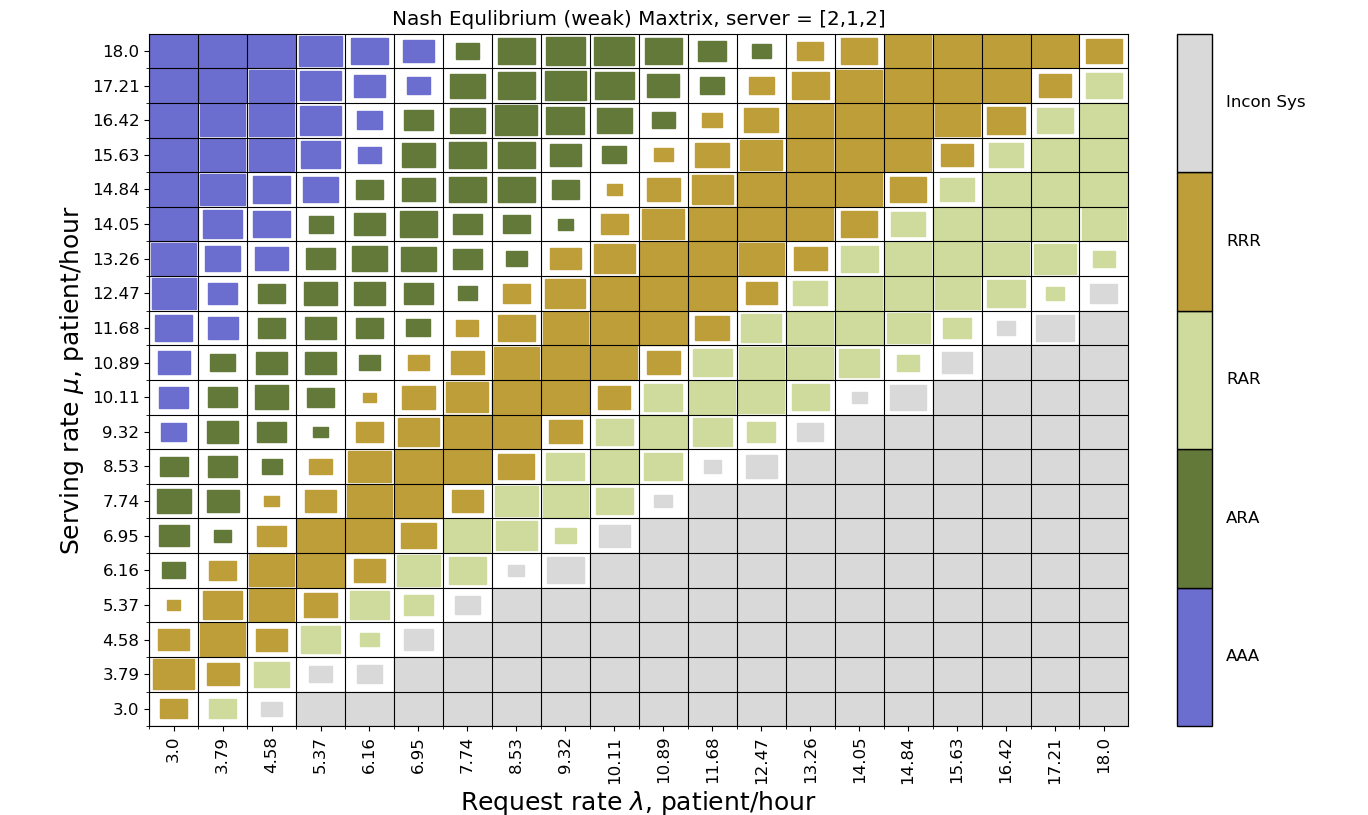}
\caption{A 2-D Nash Equilibrium matrix with the outcome argument "Score" as the payoff with the servers combination $N=[2,2,2]$ (top) and $N=[2,1,2]$ (bottom)}
\label{fig:2dNash222}
\end{minipage}\hfill
\end{figure}

\textbf{Results} 

An example of simulation results is shown in Figure~\ref{fig:2dNash222}. The figure depicts the predicted strategies with respect to the arrival rate and the global service rate. It shows a 2D Nash Equilibrium with the outcome “Score” as the payoff in the scenario of $N=[2,2,2]$ (top) and $N=[2,1,2]$ (bottom). By running 106 times simulation for each block (cell), the most possible strategy combination (Nash Equilibrium) is displayed in different colours on the inner square. The area of the inner square ratio to the area of the outer square is the probability of the occurrences. For example, the probability of AAA strategy at the left-top corner is 100\%. The figure shows that with a growing arrival rate ($\lambda$) and declining service rate ($M$), the pure Nash Equilibrium transits from AAA to RRR until the system is inconsistent. The term "inconsistent" (abbr. "Incon Sys", shown as grey square in the plot) states there is no Nash Equilibrium existed due to the system being overcrowded (meaning that queuing time for a new patient is infinite). 

    
\section{Case study: modelling ambulance dispatching for the \acrshort{acs} patients}
\label{case_study}

The \acrfull{acs} is a syndrome where the patient's heart may not be functioning properly due to decreased blood flow in the coronary arteries \cite{amsterdam20142014}. The \acrshort{acs} usually occurs unexpectedly and results in a high mortality rate. An imbalanced distribution of medical resources which results in overcrowding in regional hospitals is considered one of the factors influencing mortality. The \acrshort{acs} patients are sensitive to time so that every extra minute impacts their health outcome and specialised aid in hospitals is needed for most of them. An efficient ambulance dispatching system can facilitate a decrease in mortality rate.

To better understand the dynamics of hospital behaviours, we adopt the \acrshort{gt-des} model to simulate the ambulance dispatching process of \acrshort{acs} patients in Saint Petersburg. In addition, we evaluate the model performance by comparing the simulated and recorded annual mortality rates of hospitals. As a result, the validated model can be used to understand the healthcare system in the city and help improving the policy making for city authorities.


\subsection{Simulation settings}

\begin{figure}[!h]
\centering
\includegraphics[width=0.5\textwidth]{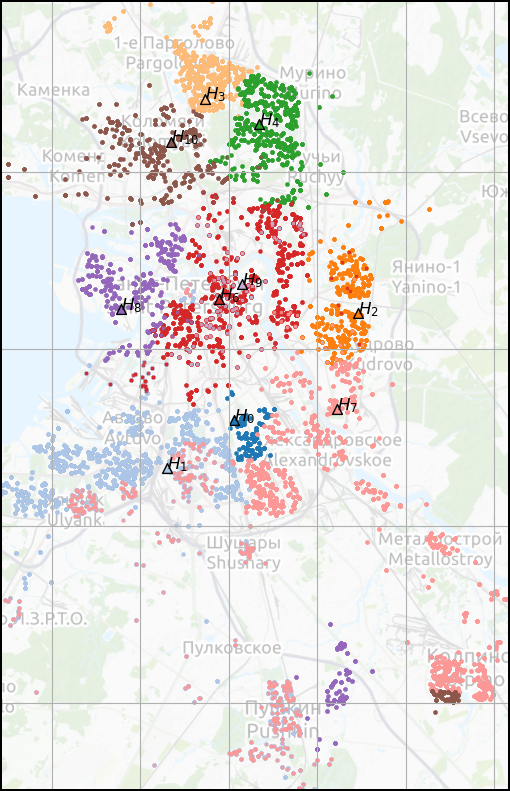}
\caption{The locations of thirteen hospitals and the corresponding visit patients in Saint Petersburg}
\label{fig:hos_loc}
\end{figure}

An spatial environment is shown in Figure~\ref{fig:hos_loc}. There are recorded 5124 \acrshort{acs} patients delivered to hospitals by the city ambulance service in Saint Petersburg from 2015-01-01 to 2015-11-15. The dataset represents approximately 20\% of all the \acrshort{acs} cases in the city during the observation time served by a centalized city ambulance service (the remaining 80\% were served by local district-level ambulances). The patients were delivered to thirteen hospitals having the \acrshort{pci} facilities denoted as $H_i$ for $i \in \{0,1,2...,12\}$. On Figure~\ref{fig:hos_loc}, $H_{11}$ and $H_{12}$ are excluded due to specific operating schedule and having a small number of visited patients. $H_5$ is excluded due to operating in a relatively remote district of the city. 

\begin{figure}[!ht]
\centering
\includegraphics[width=0.47\textwidth]{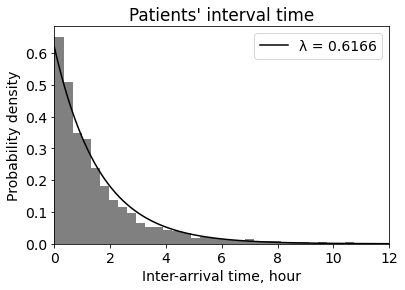}
\includegraphics[width=0.47\textwidth]{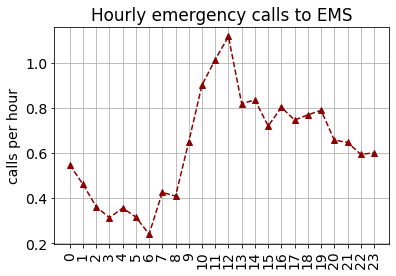}
\caption{Fitted inter-arrival time of the \acrshort{acs} patients by exponential distribution (left) and average number of requests per hour in a day (right)} 
\label{fig:fittedExp}
\end{figure}

Previously in the two-dimensional artificial setting, the inter-arrival time is assumed to have exponential distribution according to queuing theory \cite{banks2005discrete}. The left plot in Figure~\ref{fig:fittedExp} verified the assumption. The left plot shows the average interval between an \acrshort{acs} patient's request to the \acrshort{ems} is $\frac{1}{\lambda} \approx 1.4470$  hours. However, in real life, the healthcare system is more complicated since the arrival rate $\lambda$ varies hourly, as shown in the right plot in Figure~\ref{fig:fittedExp}. Consequently, the simulated arrival rate is scaled by the rush-hour effect where $\{\lambda_h\} = \lambda\ s_h$ for hour $h \in \{0,1,2...,23\}$ and $s_h$ is the scale factor at hour $h$. And to avoid geographical bias, the locations of simulated patients are sampled from recorded data.

Patient travel time is estimated by using the traffic data in the study \cite{derevitskiy2017iccs}. The estimation takes into account the dynamics of the city environment (including rush hours, weekends, and holidays). The traffic data used is represented by a directed travel network containing travel time records linking 143 departure nodes distributed across the city to 19 destination nodes (including the hospitals).

\begin{figure}[!h]
\centering
\includegraphics[width=0.43\columnwidth]{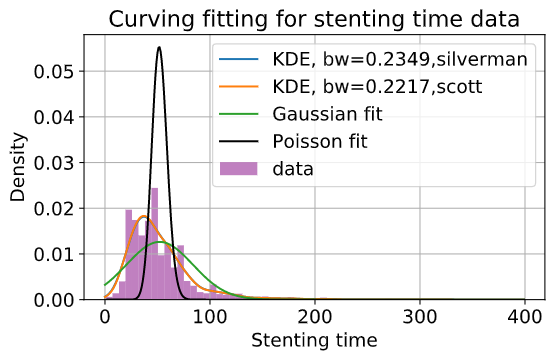}
\caption{The fitted stent time data (service time) using the \acrfull{kde} method}
\label{fig:ave_traTime}
\end{figure}

To estimate the service (\acrshort{pci}) time, a dataset containing 1866 records of stent time is analysed (the dataset is obtained from electronic health records of anonymous patients treated in the Almazov National Medical Research Centre, Saint Petersburg, Russia). Based on the data, the assumption of service time have a Poisson distribution is not true. Thus we fit the data with the underlying distribution derived from the \acrfull{kde} method. the \acrshort{kde} is used to estimate the probability density function of a variable $x$ for unequally weighted data points with respect to unimodality \cite{davis2011remarks}, we obtain the expression:

$${\displaystyle {\widehat {f}}_{h}(x)={\frac {1}{n}}\sum _{i=1}^{n}K_{h}(x-x_{i})={\frac {1}{nh}}\sum _{i=1}^{n}K{\Big (}{\frac {x-x_{i}}{h}}{\Big )}}$$

where $K$ is a selected kernel function (Gaussian function in this case because as the volume of data grows, the selection of kernel function does less impact on estimated density) under the assumptions of 1. non-negative $K(x)\ge 0$ for every $x$ 2. symmetric: $K(x)= K(-x)$. 

To validate the goodness of the \acrshort{kde} fit, the \acrfull{kstest} is utilised. The definition of the \acrshort{kstest} is that for an observed random variable $x$ where $G(x)$ is the cumulative results and $F(x)$ is a given Cumulative Distribution Function (CDF). The KStest hypothesis that the two distributions are identical as $G(x)=F(x)$. Then the CDF of the \acrshort{kde} function is 

$${\displaystyle {\widehat {F}}_{h}(x)={\frac {1}{nh}}\sum _{i=1}^{n}F_K{\Big (}{\frac {x-x_{i}}{h}}{\Big )}},$$

where $F_K$ is the CDF of the kernel function. As a result with $N=20$, the \acrshort{kstest} returns $statistic=0.16, P_{value}=0.61$ where given the 5\% significant level, we do not reject the hypothesis that the two distributions are identical as $G(x)=F(x)$ as $p_{value}>0.05$. 

\subsection{Simulation results}

The simulation aims to find out the primary strategy of each hospital and validate strategies by real-world data. The primary strategy of a hospital is generally categorised as "Accepting" and "Redirecting" which is defined as the hospital’s most likely response to any simulated patient request. Firstly, we divided all hospitals into repeatable matching pairs (i.e $ \{H_i,H_j\}$ for $i,j \in \{0,1,2,...10\}$ and $i \ne j$). Secondly, we filter the pairs with the proportion of patients two hospitals are sharing with each other. The reason is that some hospital pairs may not have interaction between each other. For instance, Figure~\ref{fig:hos_loc} shows $H_3$ and $H_1$ are so far away from each other and they only share $3\%$ patients based on the data. Next, we apply the \acrshort{gt-des} model for each filtered pair of hospitals and predict their equilibrium pattern for each hospital (similar to Figure~\ref{2dmodel}). Lastly, we estimate the primary strategy of each hospital by computing the contribution of paired strategies.  

\begin{figure}[h!]
\begin{center}
\includegraphics[width=0.85\columnwidth]{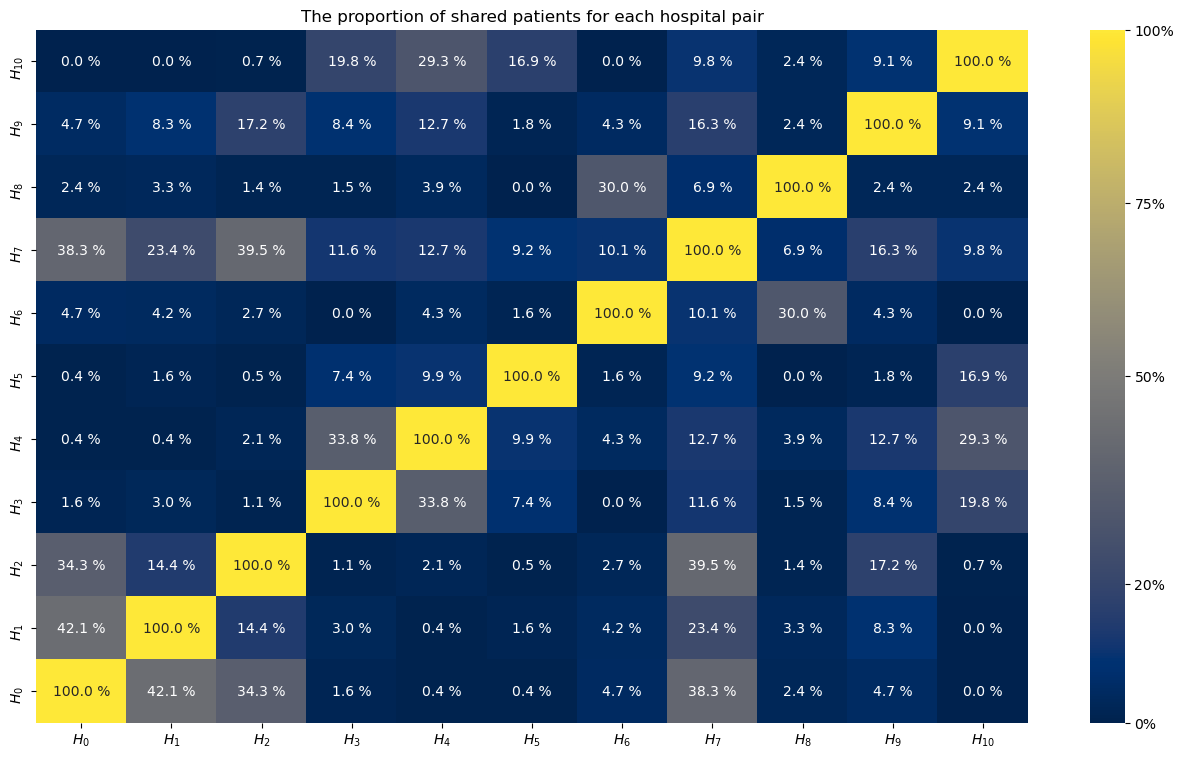}
\caption{The proportion of shared patients for each hospital pair. }
\label{fig:heat}
\end{center}
\end{figure}

We utilised spatial method to filter hospital pair by using the proportion of shared patients as the threshold. As shown in Figure~\ref{fig:hos_loc}, we spatially divide the map. Next, we compute the the proportion of shared patient for each hospital pair as shown in the Figure~\ref{fig:heat}. With the threshold being 10\%, seventeen hospital pairs are filtered. For more information, the simulation result for each filtered pair is in Appendix \ref{Appendix:hospairs}.

\begin{table}[]
\centering
\caption{Predicted strategies for each filtered hospital pairs}
\scalebox{0.8}{
\normalsize
\begin{tabular}{|c|c|l|c|l|c|l|c|l|c|l|c|l|c|l|c|l|c|l|c|l|c|l|c|}
\hline

\backslashbox{$H_i$}{$H_j$} & \multicolumn{2}{c|}{$H_0$}                                             
                                            & \multicolumn{2}{c|}{$H_1$}                                             & \multicolumn{2}{c|}{$H_2$}                                             & \multicolumn{2}{c|}{$H_3$}                                                     & \multicolumn{2}{c|}{$H_4$}                                                     & \multicolumn{2}{c|}{$H_5$}                     & \multicolumn{2}{c|}{$H_6$}                                             & \multicolumn{2}{c|}{$H_7$}                                                     & \multicolumn{2}{c|}{$H_8$}                                                    & \multicolumn{2}{c|}{$H_9$}                                                     & \multicolumn{2}{c|}{$H_{10}$}                                                  & \textbf{\begin{tabular}[c]{@{}c@{}}Weighted\\ Strategy\end{tabular}} \\ \hline
$H_0$       & \multicolumn{2}{c|}{\textbf{\textbackslash{}}}                         & \multicolumn{2}{c|}{A}                                                 & \multicolumn{2}{c|}{\begin{tabular}[c]{@{}c@{}}A\\ 0.343\end{tabular}} & \multicolumn{2}{c|}{\textbf{\textbackslash{}}}                                 & \multicolumn{2}{c|}{\textbf{\textbackslash{}}}                                 & \multicolumn{2}{c|}{\textbf{\textbackslash{}}} & \multicolumn{2}{c|}{\textbf{\textbackslash{}}}                         & \multicolumn{2}{c|}{\begin{tabular}[c]{@{}c@{}}A\\ 0.383\end{tabular}}         & \multicolumn{2}{c|}{\textbf{\textbackslash{}}}                                & \multicolumn{2}{c|}{\textbf{\textbackslash{}}}                                 & \multicolumn{2}{c|}{\textbf{\textbackslash{}}}                                 & \textbf{A}                                                           \\ \hline
$H_1$       & \multicolumn{2}{c|}{\begin{tabular}[c]{@{}c@{}}R\\ 0.421\end{tabular}} & \multicolumn{2}{c|}{\textbf{\textbackslash{}}}                         & \multicolumn{2}{c|}{\begin{tabular}[c]{@{}c@{}}R\\ 0.144\end{tabular}} & \multicolumn{2}{c|}{\textbf{\textbackslash{}}}                                 & \multicolumn{2}{c|}{\textbf{\textbackslash{}}}                                 & \multicolumn{2}{c|}{\textbf{\textbackslash{}}} & \multicolumn{2}{c|}{\textbf{\textbackslash{}}}                         & \multicolumn{2}{c|}{\begin{tabular}[c]{@{}c@{}}R\\ 0.234\end{tabular}}         & \multicolumn{2}{c|}{\textbf{\textbackslash{}}}                                & \multicolumn{2}{c|}{\textbf{\textbackslash{}}}                                 & \multicolumn{2}{c|}{\textbf{\textbackslash{}}}                                 & \textbf{R}                                                           \\ \hline
$H_2$       & \multicolumn{2}{c|}{\begin{tabular}[c]{@{}c@{}}R\\ 0.343\end{tabular}} & \multicolumn{2}{c|}{\begin{tabular}[c]{@{}c@{}}R\\ 0.144\end{tabular}} & \multicolumn{2}{c|}{\textbf{\textbackslash{}}}                         & \multicolumn{2}{c|}{\textbf{\textbackslash{}}}                                 & \multicolumn{2}{c|}{\textbf{\textbackslash{}}}                                 & \multicolumn{2}{c|}{\textbf{\textbackslash{}}} & \multicolumn{2}{c|}{\textbf{\textbackslash{}}}                         & \multicolumn{2}{c|}{\begin{tabular}[c]{@{}c@{}}A\\ 0.395\end{tabular}}         & \multicolumn{2}{c|}{\textbf{\textbackslash{}}}                                & \multicolumn{2}{c|}{\begin{tabular}[c]{@{}c@{}}0.2A:0.8R\\ 0.172\end{tabular}} & \multicolumn{2}{c|}{\textbf{\textbackslash{}}}                                 & \textbf{R(A)}                                                        \\ \hline
$H_3$       & \multicolumn{2}{c|}{\textbf{\textbackslash{}}}                         & \multicolumn{2}{c|}{\textbf{\textbackslash{}}}                         & \multicolumn{2}{c|}{\textbf{\textbackslash{}}}                         & \multicolumn{2}{c|}{\textbf{\textbackslash{}}}                                 & \multicolumn{2}{c|}{\begin{tabular}[c]{@{}c@{}}A\\ 0.338\end{tabular}}         & \multicolumn{2}{c|}{\textbf{\textbackslash{}}} & \multicolumn{2}{c|}{\textbf{\textbackslash{}}}                         & \multicolumn{2}{c|}{\begin{tabular}[c]{@{}c@{}}A\\ 0.116\end{tabular}}         & \multicolumn{2}{c|}{\textbf{\textbackslash{}}}                                & \multicolumn{2}{c|}{\textbf{\textbackslash{}}}                                 & \multicolumn{2}{c|}{\begin{tabular}[c]{@{}c@{}}A\\ 0.198\end{tabular}}         & \textbf{A}                                                           \\ \hline
$H_4$       & \multicolumn{2}{c|}{\textbf{\textbackslash{}}}                         & \multicolumn{2}{c|}{\textbf{\textbackslash{}}}                         & \multicolumn{2}{c|}{\textbf{\textbackslash{}}}                         & \multicolumn{2}{c|}{\begin{tabular}[c]{@{}c@{}}0.1A:0.9R\\ 0.338\end{tabular}} & \multicolumn{2}{c|}{\textbf{\textbackslash{}}}                                 & \multicolumn{2}{c|}{\textbf{\textbackslash{}}} & \multicolumn{2}{c|}{\textbf{\textbackslash{}}}                         & \multicolumn{2}{c|}{\begin{tabular}[c]{@{}c@{}}0.9A:0.1R\\ 0.127\end{tabular}} & \multicolumn{2}{c|}{\textbf{\textbackslash{}}}                                & \multicolumn{2}{c|}{\begin{tabular}[c]{@{}c@{}}A\\ 0.127\end{tabular}}         & \multicolumn{2}{c|}{\begin{tabular}[c]{@{}c@{}}0.8A:0.2R\\ 0.293\end{tabular}} & \textbf{A(R)}                                                        \\ \hline
$H_6$       & \multicolumn{2}{c|}{\textbf{\textbackslash{}}}                         & \multicolumn{2}{c|}{\textbf{\textbackslash{}}}                         & \multicolumn{2}{c|}{\textbf{\textbackslash{}}}                         & \multicolumn{2}{c|}{\textbf{\textbackslash{}}}                                 & \multicolumn{2}{c|}{\textbf{\textbackslash{}}}                                 & \multicolumn{2}{c|}{\textbf{\textbackslash{}}} & \multicolumn{2}{c|}{\textbf{\textbackslash{}}}                         & \multicolumn{2}{c|}{\begin{tabular}[c]{@{}c@{}}0.3A:0.7R\\ 0.101\end{tabular}} & \multicolumn{2}{c|}{\begin{tabular}[c]{@{}c@{}}0.3A:0.7R\\ 0.30\end{tabular}} & \multicolumn{2}{c|}{\textbf{\textbackslash{}}}                                 & \multicolumn{2}{c|}{\textbf{\textbackslash{}}}                                 & \textbf{R}                                                           \\ \hline
$H_7$       & \multicolumn{2}{c|}{\begin{tabular}[c]{@{}c@{}}R\\ 0.383\end{tabular}} & \multicolumn{2}{c|}{\begin{tabular}[c]{@{}c@{}}R\\ 0.234\end{tabular}} & \multicolumn{2}{c|}{\begin{tabular}[c]{@{}c@{}}R\\ 0.395\end{tabular}} & \multicolumn{2}{c|}{\begin{tabular}[c]{@{}c@{}}0.3A:0.7R\\ 0.116\end{tabular}} & \multicolumn{2}{c|}{\begin{tabular}[c]{@{}c@{}}0.7A:0.3R\\ 0.127\end{tabular}} & \multicolumn{2}{c|}{\textbf{\textbackslash{}}} & \multicolumn{2}{c|}{\begin{tabular}[c]{@{}c@{}}A\\ 0.101\end{tabular}} & \multicolumn{2}{c|}{\textbf{\textbackslash{}}}                                 & \multicolumn{2}{c|}{\textbf{\textbackslash{}}}                                & \multicolumn{2}{c|}{\begin{tabular}[c]{@{}c@{}}0.9A:0.1R\\ 0.163\end{tabular}} & \multicolumn{2}{c|}{\textbf{\textbackslash{}}}                                 & \textbf{R}                                                           \\ \hline
$H_8$       & \multicolumn{2}{c|}{\textbf{\textbackslash{}}}                         & \multicolumn{2}{c|}{\textbf{\textbackslash{}}}                         & \multicolumn{2}{c|}{\textbf{\textbackslash{}}}                         & \multicolumn{2}{c|}{\textbf{\textbackslash{}}}                                 & \multicolumn{2}{c|}{\textbf{\textbackslash{}}}                                 & \multicolumn{2}{c|}{\textbf{\textbackslash{}}} & \multicolumn{2}{c|}{\begin{tabular}[c]{@{}c@{}}A\\ 0.300\end{tabular}} & \multicolumn{2}{c|}{\textbf{\textbackslash{}}}                                 & \multicolumn{2}{c|}{\textbf{\textbackslash{}}}                                & \multicolumn{2}{c|}{\textbf{\textbackslash{}}}                                 & \multicolumn{2}{c|}{\textbf{\textbackslash{}}}                                 & \textbf{A}                                                           \\ \hline
$H_9$       & \multicolumn{2}{c|}{\textbf{\textbackslash{}}}                         & \multicolumn{2}{c|}{\textbf{\textbackslash{}}}                         & \multicolumn{2}{c|}{\begin{tabular}[c]{@{}c@{}}A\\ 0.172\end{tabular}} & \multicolumn{2}{c|}{\textbf{\textbackslash{}}}                                 & \multicolumn{2}{c|}{\begin{tabular}[c]{@{}c@{}}A\\ 0.127\end{tabular}}         & \multicolumn{2}{c|}{\textbf{\textbackslash{}}} & \multicolumn{2}{c|}{\textbf{\textbackslash{}}}                         & \multicolumn{2}{c|}{\begin{tabular}[c]{@{}c@{}}A\\ 0.163\end{tabular}}         & \multicolumn{2}{c|}{\textbf{\textbackslash{}}}                                & \multicolumn{2}{c|}{\textbf{\textbackslash{}}}                                 & \multicolumn{2}{c|}{\textbf{\textbackslash{}}}                                 & \textbf{A}                                                           \\ \hline
$H_{10}$    & \multicolumn{2}{c|}{\textbf{\textbackslash{}}}                         & \multicolumn{2}{c|}{\textbf{\textbackslash{}}}                         & \multicolumn{2}{c|}{\textbf{\textbackslash{}}}                         & \multicolumn{2}{c|}{\begin{tabular}[c]{@{}c@{}}A\\ 0.198\end{tabular}}         & \multicolumn{2}{c|}{\begin{tabular}[c]{@{}c@{}}A\\ 0.293\end{tabular}}         & \multicolumn{2}{c|}{\textbf{\textbackslash{}}} & \multicolumn{2}{c|}{\textbf{\textbackslash{}}}                         & \multicolumn{2}{c|}{\textbf{\textbackslash{}}}                                 & \multicolumn{2}{c|}{\textbf{\textbackslash{}}}                                & \multicolumn{2}{c|}{\textbf{\textbackslash{}}}                                 & \multicolumn{2}{c|}{\textbf{\textbackslash{}}}                                 & \textbf{A}                                                           \\ \hline
\end{tabular}

}
\label{tab:weightstrategy}
\end{table}

Next, we apply the \acrshort{gt-des} model to each hospital pairs to derive the strategy for each hospital. Table \ref{tab:weightstrategy} shows the derived strategy $a_{i}^{i,j}$ of hospital $H_i$ when interacting with paired hospital $H_j$. For each cell, the top half part is the predicted strategy $a_{i}^{i,j}$, and beneath it is the proportion of shared patients out of all the patients visited both hospitals $\omega_{i,j}$ (if the proportion is 100\%, the weight is not displayed). The proportion of shared patients $\omega_{i,j}$ is also the weight which is derived from Figure~\ref{fig:heat}. The weight is the coefficient of each paired strategy contributing to the primary strategy $a_i$. Consequently, the weighted strategy $a_i$ (previously called primary strategy, since weights were not involved) of $H_i$ is calculated by expression:

\begin{equation}
\begin{split}
        a_i =\argmax_{a\in\{A,R\}}( & \sum_{j=0,j\ne 5,j\ne i}^{10} \omega_{i,j} a_{i}^{i,j}; a\in\{A\},\\
        & \sum_{j=0,j\ne 5,j\ne i}^{10} \omega_{i,j} a_{i}^{i,j}; a\in\{R\}) 
\end{split}
\end{equation}

For example, to compute the weighted strategy of $H_6$ with $i=6$, the interactions between $H_6$ and $H_7,H_8$ are accounted. Thus, the cumulative chance of $a_6 = A$ is $$\sum_{j=0,j\ne 5,j\ne 6}^{10} \omega_{6,j} a_{6}^{6,j} = \omega_{6,7} a_6^{6,7} + \omega_{6,8}a_6^{6,8}= 0.101*0.3 + 0.30*0.3 = 0.1203; a\in\{A\}.$$

In addition, the cumulative chance of $a_6 = R$ is $$\sum_{j=0,j\ne 5,j\ne 6}^{10} \omega_{i,j} a_{i}^{i,j} = \omega_{6,7} a_6^{6,7} + \omega_{6,8}a_6^{6,8} = 0.101*0.7+0.3*0.7 = 0.2807; a\in\{R\}.$$

With $0.2807 >  0.1203$, the weighted strategy of $H_6$ is $a_6 = R$.

\begin{figure}[!h]
\begin{center}
\includegraphics[width=0.375\columnwidth]{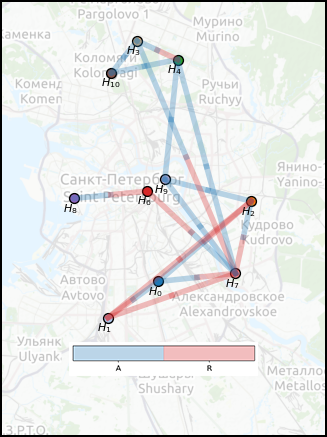}
\includegraphics[width=0.575\columnwidth]{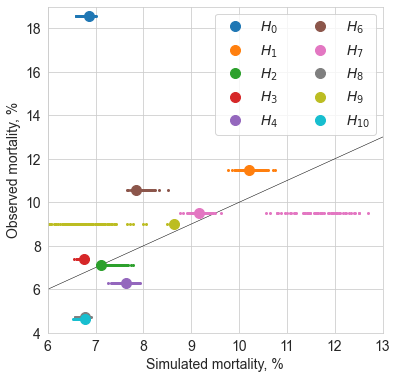}
\end{center}
\centering
\caption{Predicted strategies for pair-to-pair hospitals (left) and quantile-to-quantile plot of simulated and observed mortality (right)}
\label{fig:shared1}
\end{figure}

However, it may happen the weights to both strategies are close such as $H_2$ and $H_4$ in the Table \ref{tab:weightstrategy}. Thus, we derived that if the proportion of weighted strategy $a_{i}$ has more than 40\% (lower bounder) probability but less than 60\% probability (upper bounder) indicating the weighted strategy, both strategy options $a\in\{A,R\}$ are assumed. Because the simulation has many assumptions that will result in vibrations in weighted strategy. Eventually, the strategy combination for the hospitals in Saint Petersburg are $\vec{a} = \{A, R, R(A),A(R),A,R,R,A,A,A\}$. Figure~\ref{fig:shared1} (left) is the graphic representative of Figure~\ref{tab:weightstrategy}, which show the interactive strategies of hospital pairs.  



\subsection{\textbf{Model validation}}

To validate the predicted strategy combination, we compared the simulated mortality of each hospital to the recorded mortality. To obtain the simulated mortality, we simulate the \acrshort{gt-des} model for \acrshort{acs} patients with 512 pre-defined strategy combinations for each hospital $\{a_0,a_1,...a_{10}\} \setminus \{a_5\}$ for $a_i \in \{A,R\}$ with excluded $a_{5}$ and fixed $a_{6} = R$. For each simulation, a list of hospital mortality values is estimated using the door-to-balloon time \cite{rathore2009association}. The door-to-balloon time is one of the most important factors in treating heart attack patients, which measures the interval between patient's arrival and the surgery (the time between the arrival of a patient with STEMI in the emergency room until the time that a balloon is inflated in the occluded, culprit coronary artery) \cite{soon2007impact}. Formally, the door-to-balloon time of a patient $i$ is $T_{i,DtB} = T_{i,queuing} + T_{i,service}$ for hospital $H_i$. To eliminate systematic biases caused by the unobserved time component, we used the mortality probability estimation function provided in \cite{rathore2009association} scaled linearly with the following coefficients estimated after simulation of all available scenarios: $p'(m)= 3.0128p(m) - 3.0560$, where $p(m)$ is approximated according to referenced paper. The door-to-balloon time is averaged by 10 simulations.  

Figure~\ref{fig:shared1} (right) shows a comparison of simulated and observed mortality using the predicted strategy combination $\vec{a} = \{A, R, R,A,A,R,R,A,A,A\}$ \cite{kovalchuk2018distributed}. The large markers show that the results obtained for the predicted combination of strategies. As it can be seen from the comparison of the mortality for the predicted strategy combination with the ranges of mortality in different simulations, most of the hospitals reach a good ($H_0$, $H_2$, $H_3$, $H_7$, $H_9$) or average ($H_1$, $H_4$) mortality prediction. Only in a few of the hospitals, the mortality is predicted as worse than average ($H_6$, $H_8$, $H_{10}$). Still, all the latter hospitals show a relatively low variation of mortality during strategy selection. The hospital $H_0$ is deviating from the linear fit because $H_0$ is a specialised stent hospital, which deals with \acrshort{acs} patients with the severe symptoms. Thus, the mortality of $H_0$ is an outlier. 

Next, we conduct a comparative analysis between the simulated and real-world mortality rate for each hospital as system-level macro-characteristics. To assess the proposed methods as a way of identification of behaviour diversity in healthcare agents, we used the Pearson correlation with the observed mortality. The mortality results using the predicted strategy have the generally Top 3 highest Pearson correlation with the observed mortality, compared to any other random strategy combination used.

\begin{table}[!h]
\centering
\caption{The Pearson correlation results using three typical strategy combinations}
\begin{tabular}{|c|c|c|}
\hline
\textbf{Strategy combination} & \textbf{Pearson correlation} & \textbf{P-value} \\
\hline
{[}'A', 'A', 'A', 'A', 'A', 'A', 'A', 'A', 'A', 'A'{]} & 0.73 & 0.0436 \\
\hline
{[}'R', 'R', 'R', 'R', 'R', 'R', 'R', 'R', 'R', 'R'{]} & 0.74 & 0.0256 \\
\hline
{[}'A', 'R', 'A', 'A', 'R', 'R', 'R', 'A', 'A', 'A'{]} & 0.80 & 0.0064 \\
\hline
\end{tabular}
\label{pearsonval}
\end{table}

\begin{figure}[h!]
\begin{center}
\includegraphics[width=0.6\columnwidth]{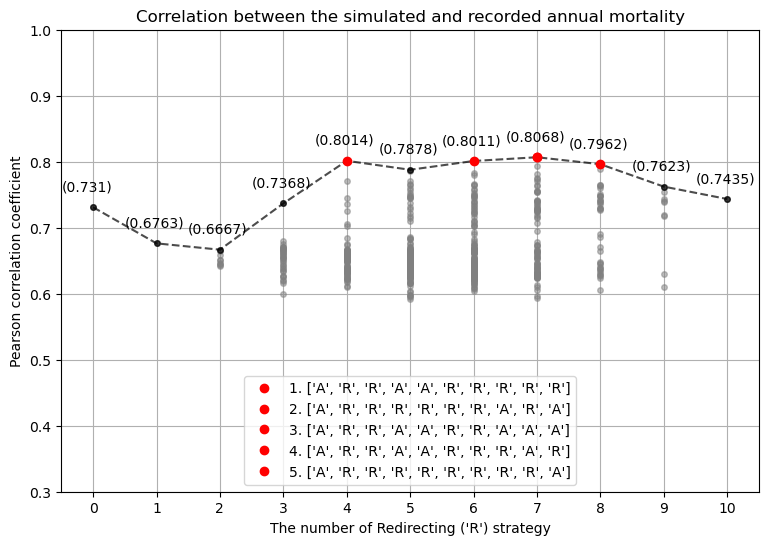}
\caption{The Pearson's r between the observed mortality and simulated mortality by using every possible strategy combination}
\label{fig:pearson}
\end{center}
\end{figure}

Table \ref{pearsonval} shows the Pearson correlation tests conducted for simulated and recorded mortality values in the situation of the “all ’A’”, “all ’R’” and the predicted strategy combination. The mortality correlation using predicted strategies being 0.8014 and the p-value being 0.0064 for a non-correlation test,  we rejected the non-correlation hypothesis by assuming the 1\% significant level. Besides, Figure~\ref{fig:pearson} the simulated mortality derived from the predicted strategy combination has high correlation (ranked top 3/512) to the recorded mortality. Due to many assumptions we made and the complications of the real system, the match-up correlation is not the highest. This will be discussed in Section \ref{discussion}. 

\section{Discussion}
\label{discussion}

The presented study is aimed towards investigating the possible ways of identifying diverse behaviour in healthcare agents within a large city. The key problem within the issue is the lack of detailed information on the behaviour of agents and the complex nature of the processes affected by multiple aspects (legal, social, economic, and even psychological). Considering the complex environment, one of the ways to investigate such a system is building a model “bottom-up” with the involvement of agent-based and game-theoretical approaches. This is very important in the case of designing a mechanism \cite{SCHWEITZER2020} aimed towards the regulation of the multi-agent environment. For example, a proper optimal (or close to optimal) regulation of ambulance dispatching through policies or financial regulation (widely discussed in the healthcare community) requires tools (a) for discovering the internal behavioural patterns and drivers; (b) for simulation-based policy optimisation. Within our study, we’ve tried to introduce a coupled (\acrshort{gt-des}) approach to discover the diversity in the behaviour of healthcare agents. Although we took a lot of assumptions on the details of process implementation and environment structure, we believe that the obtained results show that taking into account the diversity of the agents’ behaviour brings an important opportunity to discover the detailed structure of a multi-agent healthcare environment with higher precision (the correlation is improved by 12–14\% as compared with heterogeneous strategy combinations) of further modelling and simulation tasks. We see that the approach may work with situations where the behaviour of a hospital within a system significantly influences the global performance characteristics. E.g., hospitals H7 and H9 with the largest range of mortality rates obtained in different simulations (see Figure~\ref{fig:shared1})) were predicted relatively well. Furthermore, an important aspect is that by applying this approach for the purposes of simulation-based policy optimisation we can improve individual behaviour of the agents in order to increase system performance (e.g. in case of ambulance dispatching for an \acrshort{acs} patient, this may decrease the average mortality rate through lower delays in patients' delivery).

Still, to be used in real-world cases, the implemented model needs to incorporate several additional aspects. First, the serving process is much more complicated and requires taking into account schedules of clinical facilities (e.g. in Saint Petersburg, several hospitals provide extra \acrshort{pci} facilities in the daytime), different internal rules on surgery preparation, physicians’ shifts, processing of regular patients, etc. Second, patient dispatching is also a complicated decision that is affected by multiple factors including the state of the hospitals and city traffic at a particular moment, patient condition and complications, organisational regulation of ambulances, etc. Third, patient inflow varies significantly both in space and time depending on time and environment. Next, the diversity in hospital behaviour may have a more complicated structure (rather than binary selection) and may be dynamic depending on hospital load, patient inflow, condition, and complications, etc. Finally, the behaviour of all the stakeholders within the system (patients, the \acrshort{ems}, ambulance team, hospitals, physicians) are affected by multiple factors (including irrational ones). We consider further investigation of the procedure of ambulance dispatching (including stakeholder interviews, collecting larger and more detailed data, incorporating more complicated models for delivery and mortality, etc.) as a future direction for the development of the presented research.


\section{Conclusion}
\label{conclusion}

Within the presented study, an approach to simulation of the multi-agent healthcare environment is presented based on coupled \acrshort{gt-des} model for discovering of diverse behaviour strategies of the agents. The approach is studied using a problem of ambulance dispatching for \acrshort{acs} patient delivery to target hospitals. The approach is elaborated through a series of cases starting from a simplified 1D model and proceeding to a more complicated 2D model and, finally, to the real-world case of the healthcare environment of the city of Saint Petersburg. The studies show that the approach allows revealing the deviation of global optimum within a non-cooperative assumption of players. Also, a real-world case study shows that the approach enables deeper and more precise identification of a healthcare agent’s behaviour. The proposed approach may be further extended and developed to be used within policymaking and decision support systems. In addition, the proposed approach shows concepts that may be applied to many real-world situations and cases. E.g., it can be applied to different distributed multi-agent services in a city in healthcare and other domains.

\section*{Acknowledgement}

This work is supported financially by the Ministry of Science and Higher Education of the Russian Federation, Agreement No. 075-15-2021-1013 (08.10.2021) (Internal project number 13.2251.21.0067).

\bibliographystyle{ieeetr}  
\bibliography{references}
\printglossary[type=\acronymtype]

\clearpage
\appendix
\counterwithin{figure}{section}
\setcounter{figure}{0}
\counterwithin{table}{section}
\setcounter{table}{0}

\section{Appendix}
\subsection{Sobol method for sensitivity analysis}
\label{Appendix:samplesize}

\begin{figure}[h!]
\centering
\includegraphics[width=0.8\textwidth]{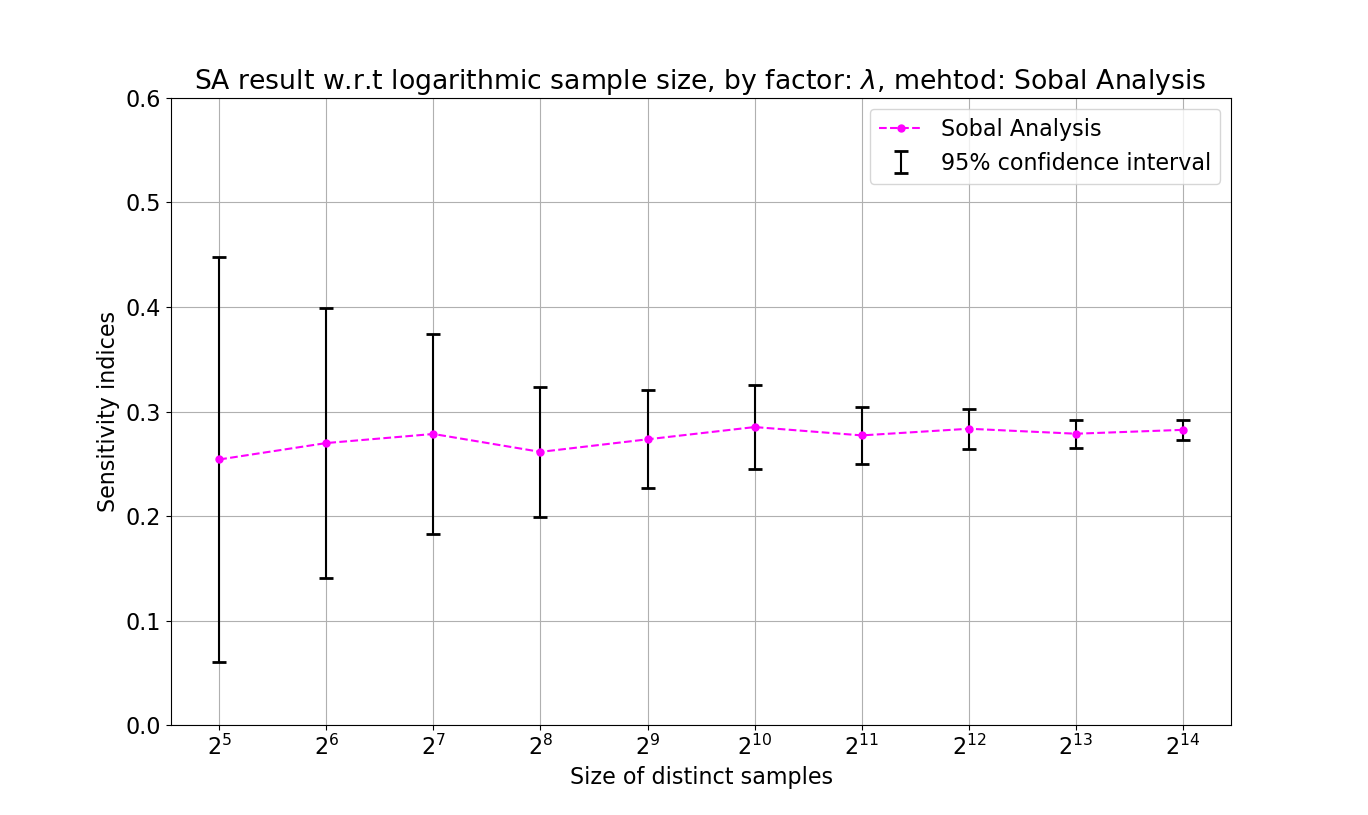}
\caption{Total-order sensitivity indies of $\lambda$ with respect to the growing size of distinct samples $N$.}
\label{fig:Sobol_size}
\end{figure}

To numerically perform the Sobol indices, we firstly use Sobol shuffle (also called as Sobol sequence) to sample the $N(d+2)$ input samples via Monte Carol approach; where the $d$ is the number of parameters ($d=9$ in the proposed model), and $N$ is the predefined size of the distinct sample. Sobol shuffle provides low-discrepancy sequences randomly, and it is an efficient sampling method. As $N$ grows, the computational cost increase linearly by O(n). To determine the size of the distinct sample, $N$, as shown in Figure \ref{fig:Sobol_size}, with the size of distinct samples increase by logarithmic, the 95\% confidence error can reach $\pm 0.1$ at $N=2^{14}$ as shown in Table \ref{tab:sa_samplesize} with two significant digits saved. Thus, the Sobol indices with $N=2^{14}$ will be performed. 

\begin{table}[!h]
\centering
\caption{SA indices of $\lambda$ w.r.t sample size}
\begin{tabular}{|l|l|l|l|l|l|l|}
\hline
\textbf{Distinct sample size}  & \textbf{$2^5$} & \textbf{$2^7$} & \textbf{$2^9$} & \textbf{$2^{11}$} & \textbf{$2^{13}$} & \textbf{$2^{14}$} \\ \hline
\textbf{SA total indices of  $\lambda$} & 0.25           & 0.27           & 0.28           & 0.28              & 0.28              & 0.28              \\ \hline
\textbf{95\% confidence error}          & $\pm0.19$           & $\pm0.09$           & $\pm0.04$           & $\pm0.02$              & $\pm0.01$              & $\pm0.01$              \\ \hline
\end{tabular}
\label{tab:sa_samplesize}
\end{table}






\subsection{Hospitals pairs selection using spacial method}
\label{Appendix:hospairs}
$[H_1, H_0],
 [H_2, H_1],
 [H_2, H_0],
 [H_4, H_3],
 [H_7, H_6],$
 
 $
 [H_7, H_4],
 [H_7, H_3],
 [H_7, H_2],
 [H_7, H_1],
 [H_7, H_0],
 [H_8, H_6],$
 
 $
 [H_9, H_7],
 [H_9, H_4],
 [H_9, H_2],
 [H_{10}, H_5],
 [H_{10}, H_4],
 [H_{10}, H_3]
 $

\begin{figure}[!h]
\begin{center}
\includegraphics[width=1\columnwidth]{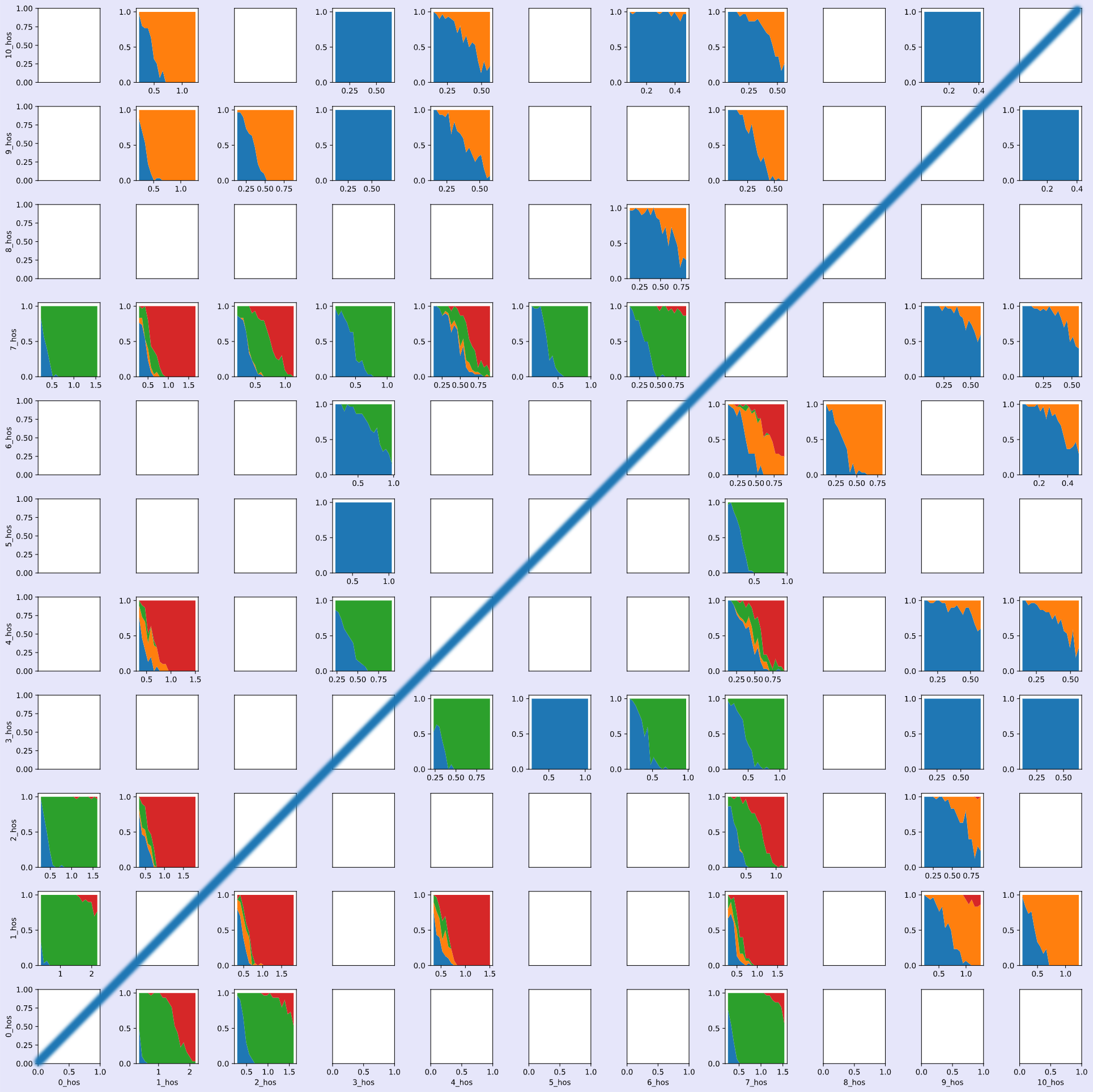}
\includegraphics[width=0.5\columnwidth]{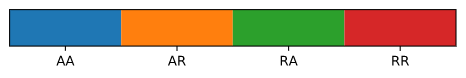}
\caption{ The simulated Nash Equilibrium for each hospital pair}
\label{fig:equmat}
\end{center}
\end{figure}

Figure \ref{fig:equmat} show the simulated Nash Equilibrium for each hospital pair (blank figures mean there is less than 10\% of shared patients between two hospitals). The equilibrium in the upper and lower triangular area considers with and without traffic jam respectively. And the equilibrium is symmetric to the diagonal line. It shows traffic jam barely influences the Nash Equilibrium. In addition, the result matches to the result of sensitivity analysis except for $H_7$ and $H_6$. And it is reasonable because a river separates the two hospitals which enhanced the influence of traffic time and the one of the bridge is not opening at night time.

\end{document}